\documentclass{article}

\usepackage[preprint]{neurips_2023}
\usepackage{longtable}
\usepackage{tabularx}
\usepackage{geometry}
\usepackage{threeparttable}
\usepackage{booktabs} 
\usepackage{dcolumn}  
\usepackage{siunitx}  
\usepackage{caption}  
\usepackage{graphicx}

\newcolumntype{d}[1]{D{.}{.}{#1}} 

\usepackage[utf8]{inputenc} 
\usepackage[T1]{fontenc}    
\usepackage{hyperref}       
\usepackage{url}            
\usepackage{booktabs}       
\usepackage{amsfonts}       
\usepackage{nicefrac}       
\usepackage{microtype}      
\usepackage{xcolor}         

\title{Forecasting Tech Sector Market Downturns based on Macroeconomic Indicators}

\author{%
  Morteza Maleki \\   
  College of Computing\\
  Georgia Institute of Technology\\
  Atlanta, Georgia\\
  mmaleki3@gatech.edu
}

\begin{document}

\maketitle

\begin{abstract}
Predicting stock price movements is a pivotal element of investment strategy, providing insights into potential trends and market volatility. This study specifically examines the predictive capacity of historical stock prices and technical indicators within the Global Industry Classification Standard (GICS) Information Technology Sector, focusing on companies established before 1980. We aim to identify patterns that precede significant, non-transient downturns—defined as declines exceeding 10\% from peak values. Utilizing a combination of machine learning techniques, including multiple regression analysis, logistic regression, random forest classification, and support vector machines, we analyze an enriched dataset comprising both macroeconomic indicators and market data. Our findings suggest that certain clusters of technical indicators, when combined with broader economic signals, offer predictive insights into forthcoming sector-specific downturns. This research not only enhances our understanding of the factors driving market dynamics in the tech sector but also provides portfolio managers and investors with a sophisticated tool for anticipating and mitigating potential losses from market downturns. Through a rigorous validation process, we demonstrate the robustness of our models, contributing to the field of financial analytics by offering a novel approach to predicting market downturns with significant implications for investment strategies and economic policy planning.
\end{abstract}

\section{Introduction}

Stock price prediction remains a fundamental challenge and opportunity in market speculation, offering insights into price trends, market volatility, and overall predictive power. Such insights are invaluable to portfolio managers and investors, guiding strategic decisions to maximize returns and mitigate risks. At the same time, with advances in technology, methods, machine learning and artificial intelligence techniques, researchers are increasingly employing them to solve real world problems \cite{maleki2022social, maleki2024clustering} and make an impact on lives of many as a result \cite{maleki2024clinical, maleki2023covid, maleki2024identification}. This work delves into the GICS Information Technology Sector, specifically analyzing companies established before 1980, to examine the predictive capacity of stock prices and technical indicators on significant, non-transient downturns—those exceeding a 10\% drop from peak values.

Market downturns manifest in various forms, each with distinct characteristics and implications for investment strategies:

\begin{itemize}
    \item Pullbacks, involving 5\% to 10\% declines, represent short-term fluctuations, typically reversing within a couple of months.
    \item Corrections reflect more substantial slides of 10\% to 20\%, marked by heightened volatility and a longer recovery period.
    \item Crashes, with downturns exceeding 20\%, can herald prolonged market depressions, extending recovery to several years.
\end{itemize}

The primary goal of this study is to innovate a predictive indicator for significant price declines, offering a strategic tool for investors to preemptively reallocate assets or capitalize on emerging opportunities. Beyond the financial markets, the repercussions of such downturns extend to government planning and international relations, underscoring the broader importance of accurate forecasts.

Recent advancements in predictive analytics and machine learning offer new perspectives on financial market dynamics. For instance, \cite{zhong2019predicting} et al. illustrate the growing sophistication of neural network models in capturing the complexities of market behavior, significantly improving the accuracy of stock price predictions. Furthermore, research by \cite{kumar2006forecasting} et al. demonstrate the potential of machine learning techniques in deciphering the intricate relationships between economic indicators and stock market performance, suggesting a paradigm shift in predictive analytics.

Moreover, the integration of alternative data sources, such as social media sentiment and blockchain activity, has been shown to enhance predictive models, offering nuanced insights into investor behavior and market trends (Smith and Kumar, 2022). These developments underscore the evolving landscape of financial analytics, where traditional and novel data sources converge within advanced computational frameworks to forecast market movements with unprecedented precision.

This study builds upon the foundational work of \cite{akaev2020forecasting}, and \cite{borio2019predicting}, incorporating the latest methodological advancements and empirical insights to refine the prediction of tech sector downturns. By navigating the intersection of economic cycles, technical indicators, and machine learning, this research aims to contribute a robust analytical tool to the arsenal of portfolio managers, enhancing their ability to navigate the volatile terrains of the tech sector.

\newpage 

\begin{longtable}{p{0.3\textwidth} p{0.2\textwidth} p{0.5\textwidth}}
\caption{List and Description of Variables Used in Analysis}\\
\hline
& &\\
\textbf{Variable} & \textbf{Category} & \textbf{Description} \\
& &\\
\hline
\endfirsthead

\multicolumn{3}{c}%
{\tablename\ \thetable\ -- \textit{Continued from previous page}} \\
\hline
\textbf{Variable} & \textbf{Category} & \textbf{Description} \\
\hline
\endhead

\hline \multicolumn{3}{r}{\textit{Continued on next page}} \\
\endfoot

\hline
\endlastfoot

Standard Deviation & Technical & Measures the amount of variability or dispersion around an average. \\
\hline
Beta & Technical & Measures the volatility of a stock or portfolio compared to the market as a whole. \\
\hline
Moving Average Convergence Divergence (MACD) & Technical & A trend-following momentum indicator that shows the relationship between two moving averages of prices. \\
\hline
Relative Strength Index (RSI) & Technical & Measures the speed and change of price movements to identify overbought or oversold conditions. \\
\hline
On-Balance Volume (OBV) & Technical & Uses volume flow to predict changes in stock price. \\
\hline
Earnings per Share (EPS) & Financial & Indicates the profitability of a company, calculated as net income divided by the number of outstanding shares. \\
\hline
Price to Earnings (P/E) Ratio & Financial & Measures a company's current share price relative to its per-share earnings. \\
\hline
Price to Earnings Ratio to Growth Ratio (PEG) & Financial & Adjusts the P/E ratio by the rate of earnings growth, providing a valuation metric that incorporates future growth. \\
\hline
Price to Book Value Ratio (P/B) & Financial & Compares a firm's market value to its book value, indicating how valued a company is in relation to its actual assets. \\
\hline
Consumer Sentiment & Macroeconomic & Gauges the overall health of the economy through consumer opinion surveys. \\
\hline
Consumer Price Index (CPI) & Macroeconomic & Measures the average change over time in the prices paid by urban consumers for a market basket of consumer goods and services. \\
\hline
Dividend Payout Ratio (DPR) & Financial & Shows the percentage of earnings distributed to shareholders in the form of dividends. \\
\hline
Dividend Yield & Financial & Measures the dividend income relative to the price of a stock. \\
\hline
Durable Goods & Macroeconomic & Indicates the value of orders placed for relatively long-lasting goods, a measure of manufacturing health. \\
\hline
Federal Funds Rate & Macroeconomic & The interest rate at which depository institutions trade federal funds with each other overnight. \\
\hline
Gross Domestic Product (GDP) & Macroeconomic & Represents the total dollar value of all goods and services produced over a specific time period. \\
\hline
Inflation & Macroeconomic & Measures the rate at which the general level of prices for goods and services is rising, and subsequently eroding purchasing power. \\
\hline
Treasury Yield & Macroeconomic & The return on investment, expressed as a percentage, on the U.S. government's debt obligations. \\
\hline
Unemployment & Macroeconomic & Represents the percentage of the total labor force that is unemployed and actively seeking employment during the specified period. \\

\end{longtable}

\newpage

\section{Material and Methods}
\subsection{Data}

The foundation of our predictive model is a robust dataset compiled from diverse and reliable sources, primarily focused on the stock market and broader economic indicators. This dataset is critical for developing and validating our forecasting model, designed to predict significant, non-transient downturns in the tech sector. Below is a detailed breakdown of the data collected:

\paragraph{Daily Stock Prices:} Utilizing the \texttt{tidyquant} library in R, with the \texttt{tq\_get()} function, we accessed Yahoo Finance to retrieve daily stock price data. This dataset encompasses:
\begin{itemize}
    \item \textbf{Ticker Symbol:} Unique identifiers for each company within the GICS Information Technology Sector, focusing on firms established before 1980.
    \item \textbf{Date:} The trading date for each stock price entry, allowing for time-series analysis.
    \item \textbf{Opening Price:} The price at which a stock first traded upon the opening of an exchange on a given trading day.
    \item \textbf{High:} The highest price at which a stock traded during the course of the trading day.
    \item \textbf{Low:} The lowest price at which a stock traded during the course of the trading day.
    \item \textbf{Closing Price:} The final price at which a stock traded during the course of the trading day.
    \item \textbf{Trading Volume:} The total number of shares or contracts traded for a particular security.
    \item \textbf{Adjusted Price:} The closing price after adjustments for all applicable splits and dividend distributions.
\end{itemize}

\paragraph{Technical Indicators:} Through Alpha Vantage, we acquired technical indicators for each listed company. These indicators, crucial for technical analysis, include:
\begin{itemize}
    \item \textbf{MACD (Moving Average Convergence Divergence):} A trend-following momentum indicator that shows the relationship between two moving averages of a security’s price.
    \item \textbf{RSI (Relative Strength Index):} A momentum oscillator that measures the speed and change of price movements, typically used to identify overbought or oversold conditions.
    \item \textbf{Standard Deviation:} A statistical measure of market volatility, indicating the dispersion of returns for a given security or market index.
\end{itemize}

\paragraph{Socioeconomic and Macroeconomic Variables:} In addition to stock prices and technical indicators, our dataset is enriched with socioeconomic and macroeconomic variables, sourced from reputable databases. These variables are integrated using the ‘Date’ variable, enabling a comprehensive analysis that factors in external influences on stock prices. These variables include, but are not limited to:
\begin{itemize}
    \item \textbf{GDP Growth Rate:} The rate at which a nation’s Gross Domestic product (GDP) changes/grows over time, indicating the economic health of the country.
    \item \textbf{Unemployment Rate:} The percentage of the labor force that is jobless and actively looking for employment.
    \item \textbf{Consumer Price Index (CPI):} A measure that examines the weighted average of prices of a basket of consumer goods and services, such as transportation, food, and medical care.
\end{itemize}

By merging these datasets on the ‘Date’ variable, we have constructed a comprehensive and aggregated dataset that serves as the backbone for our predictive model. The inclusion of both market-specific data and broader economic indicators allows for a multifaceted analysis, enhancing our ability to accurately predict significant downturns in the tech sector’s stock prices. This enriched dataset not only facilitates the exploration of direct relationships between stock prices and technical indicators but also enables the examination of how broader economic conditions impact market performance.

\noindent 
\begin{tabularx}{\textwidth}{@{}XXX@{}} 

\begin{tabular}{@{}ll@{}}
\textbf{Ticker} & \textbf{Company} \\
\hline
GLW & Corning \\
IBM & IBM \\
MSI & Motorola Solutions \\
TXN & Texas Instruments \\
APH & Amphenol \\
HPQ & HP \\
ADP & ADP \\
NXPI & NXP \\
IT & Gartner \\
STX & Seagate \\
\end{tabular}
&
\begin{tabular}{@{}ll@{}}
\textbf{Ticker} & \textbf{Company} \\
\hline
V & Visa \\
TER & Teradyne \\
AVGO & Broadcom \\
BR & Broadridge \\
ADI & Analog Devices \\
MA & Mastercard \\
TYL & Tyler Technologies \\
MU & Micron \\
TRMB & Trimble \\
ORCL & Oracle \\
\end{tabular}
&
\begin{tabular}{@{}ll@{}}
\textbf{Ticker} & \textbf{Company} \\
\hline
AMAT & Applied Materials \\
FIS & FIS \\
ANSS & Ansys \\
WDC & Western Digital \\
PAYX & Paychex \\
MSFT & Microsoft \\
KLAC & KLA \\
JKHY & Jack Henry \& Associates \\
AAPL & Apple \\
\end{tabular}
\end{tabularx}

\subsection{Variables}

This section outlines the variables considered in our analysis, aiming to determine their combined effect on market behavior, particularly in identifying significant non-transient downturns and informing buy/sell decisions. Our investigation revolves around several key inquiries:

\begin{enumerate}
    \item Identification of variable combinations predictive of significant non-transient downturns during major economic events (e.g., dot-com bubble, great recession, COVID-19 pandemic).
    \item Exploration of whether k-means clustering of non-returns based indicators can identify stocks poised to outperform the market.
    \item Analysis of whether key indicators within one set of stocks mirror those across the broader sector.
    \item Examination of active vs. passive trading strategies for retirement savings, including the comparative benefits of Mutual Funds vs. ETF investments.
\end{enumerate}

\paragraph{List of Key Variables:}
\begin{enumerate}
    \item \textbf{Relative Strength Index (RSI):} Calculated over a default period of 14 days, RSI values range from 0 to 100. Low RSI levels (below 30) suggest oversold conditions, potentially indicating a buy signal, while high RSI levels (above 70) suggest overbought conditions, potentially indicating a sell signal.
    \item \textbf{Earnings per Share (EPS):} Represents a company's profit divided by the outstanding shares of its common stock, serving as a marker of profitability. EPS can be adjusted for extraordinary items and potential share dilution.
    \item \textbf{On-Balance Volume (OBV):} A momentum indicator using volume flow to predict changes in stock price, signifying the trading volume's contribution to price movements.
    \item \textbf{Price to Earnings (P/E) Ratio:} Measures a company's current share price relative to its EPS. The P/E ratio, also known as the price multiple or earnings multiple, provides insights into stock valuation relative to earnings.
    \item \textbf{Price to Earnings Growth (PEG) Ratio:} A stock's P/E ratio divided by the growth rate of its earnings for a specified time period, offering a valuation metric that considers future earnings growth.
    \item \textbf{Price to Book Value (P/B) Ratio:} Compares a company's market price to its book value, traditionally favoring values under 1.0 as potentially undervalued stocks. However, industry variations may influence the interpretation of "good" P/B values.
    \item \textbf{Dividend Payout Ratio (DPR):} The percentage of earnings paid to shareholders in dividends, indicating how much income is distributed versus retained for growth or debt repayment.
    \item \textbf{Dividend Yield:} Reflects the dividend income relative to the stock price, offering insights into the income generated from an investment relative to its price.
\end{enumerate}

These variables are derived from the aggregated dataset comprising daily stock prices, technical indicators, and macroeconomic variables, and are pivotal in our analysis for predictive modeling. Through comprehensive evaluation of these variables, the study seeks to establish a correlation with significant market movements, particularly non-transient downturns, thereby informing strategic investment decisions.

\begin{figure*}
    \centering
    \includegraphics[width=12cm]{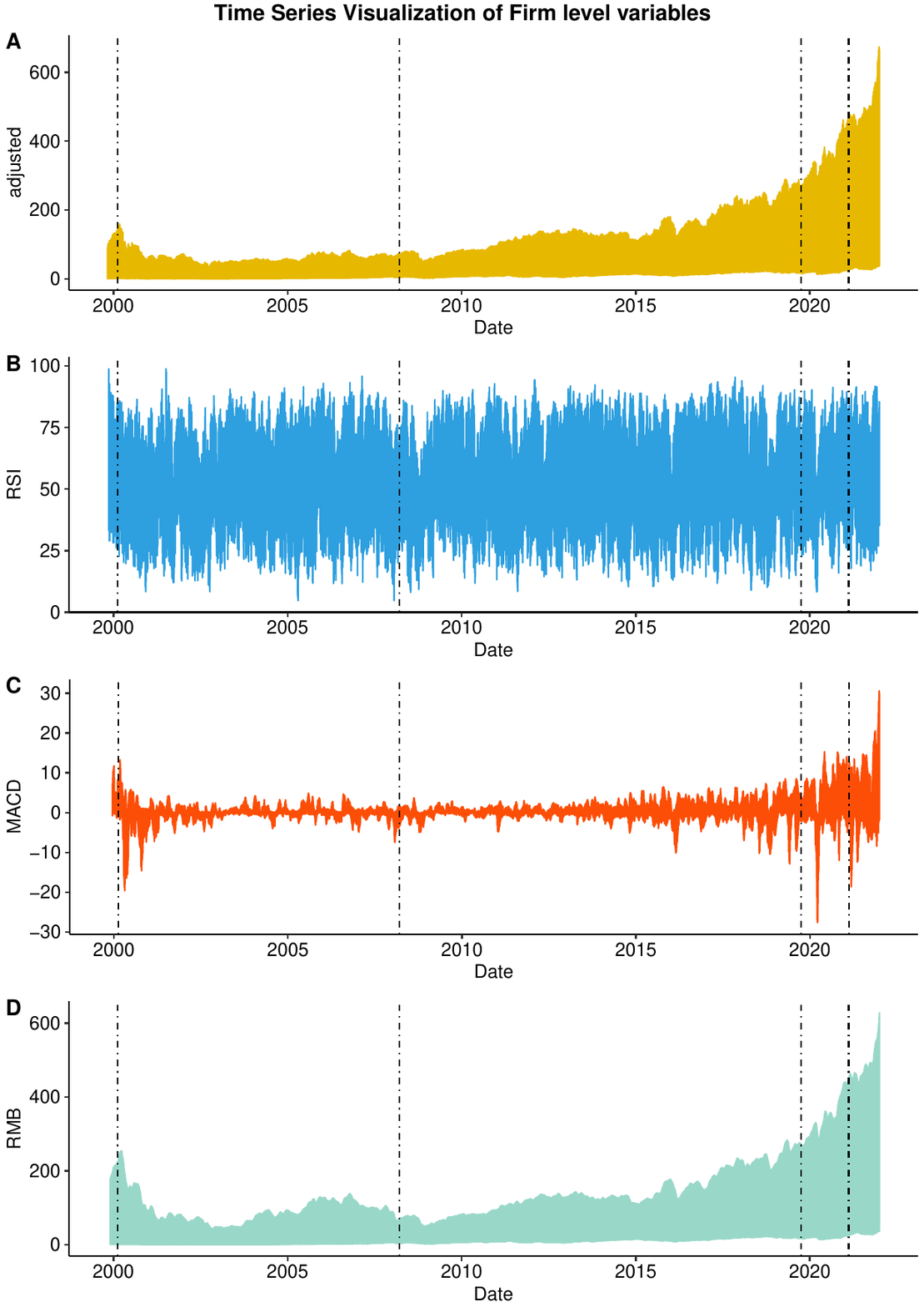}
    \caption{Time series valuation of aggregate firm level indicators (2000-2022)}
    \label{fig:enter-label}
\end{figure*}

\subsection{Methodology}
After the compilation of our datasets, a rigorous process of data cleaning, merging, and preparation was undertaken to ensure the data's readiness for further analysis. The `Date` variable served as a key axis for merging different technical indicators with the pricing data. To facilitate collaborative adjustments and enhancements, all processed data and associated code were uploaded to GitHub, allowing seamless access for all team members.

The initial datasets lacked explicit markers for the onset and conclusion of drawdown periods. To address this gap, we introduced categorizations for price drops exceeding 5\% (dips), 10\% (corrections), and 20\% (crashes), along with annotations for the troughs—representing the lowest prices within these drawdown intervals. This classification system was pivotal in refining our validation mechanisms.

Preliminary exploratory analysis was employed to dissect the available data, seeking out initial patterns worthy of deeper investigation. Correlation analyses revealed expected and unexpected relationships among the variables; for instance, a perfect negative correlation between 30-year yields and the Consumer Price Index (CPI) contrasted with an anticipated but absent correlation with inflation. Moreover, the analysis indicated that treasury yields were positively interrelated, yet disconnected from consumer sentiment metrics.

Further exploratory endeavors, as illustrated in designated figures, highlighted the Nasdaq's pronounced price variance relative to the Russell 3000 and S\&P 500 indices. This variance, significantly skewed by outliers, suggests potential investment opportunities by marking deviations far from the market baseline.

\begin{figure*}
    \centering
    \includegraphics[width=14cm]{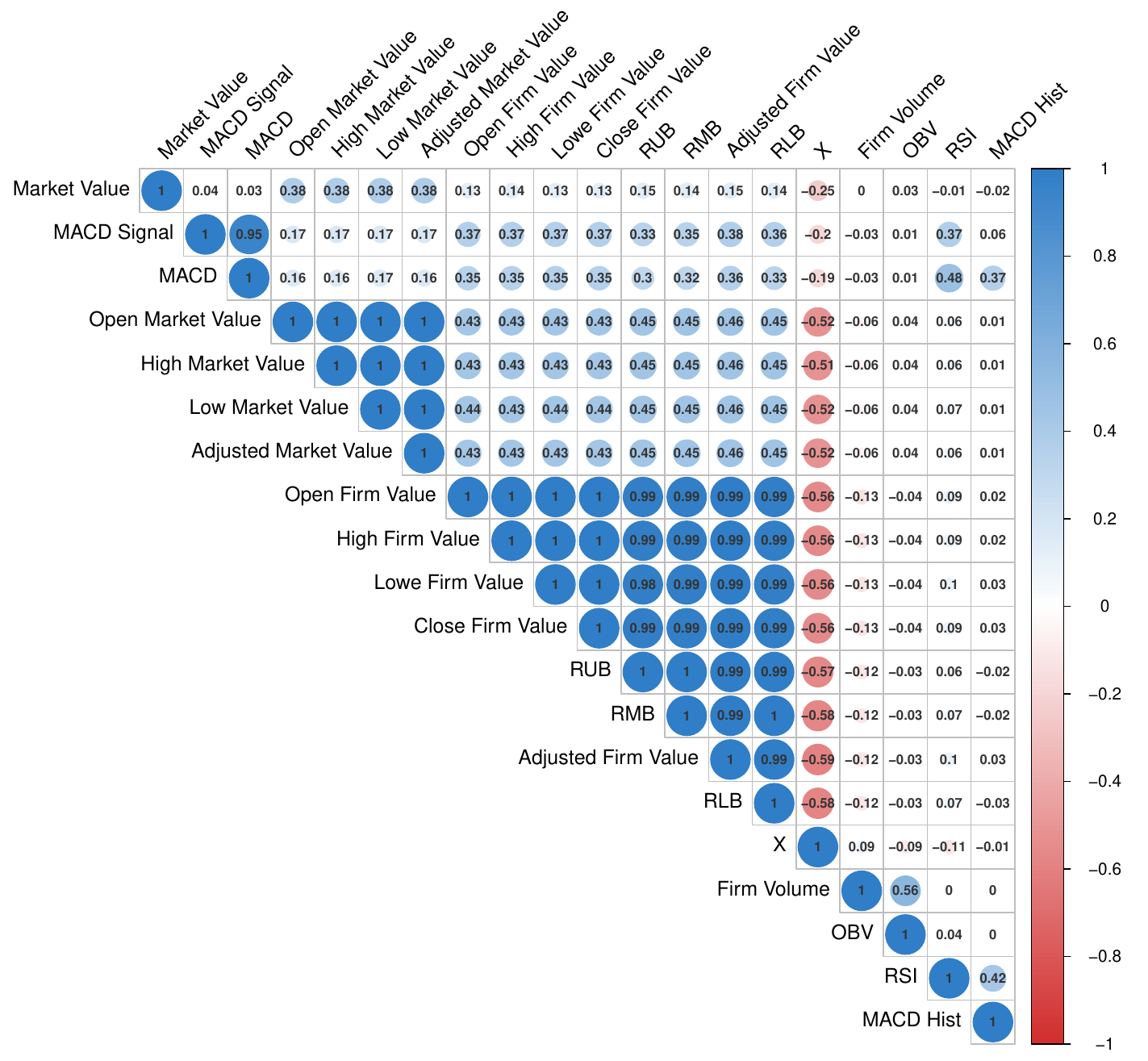}
    \caption{Aggregate market value by market sector (2000-2022)}
    \label{fig:enter-label}
\end{figure*}

Incorporating insights from preceding recession studies, such as the "Buffet Indicator" and the impact of macroeconomic variables like GDP, credit levels, and unemployment rates, we sourced additional data from AlphaVantage. Although not every factor was expected to enhance our model's predictive capability, each was evaluated for its potential contribution, with non-contributory elements being phased out.

Our proposed solution to the identified problem involves examining correlations among subgroups of stock prices and their technical indicators. Through the application of K-Means clustering to various price and technical indicators, we aim to achieve unsupervised classification of companies exhibiting similar characteristics. The optimal cluster count for each indicator set will be determined using the elbow method, laying a groundwork for predictive modeling.

\begin{figure*}
    \centering
    \includegraphics[width=14cm]{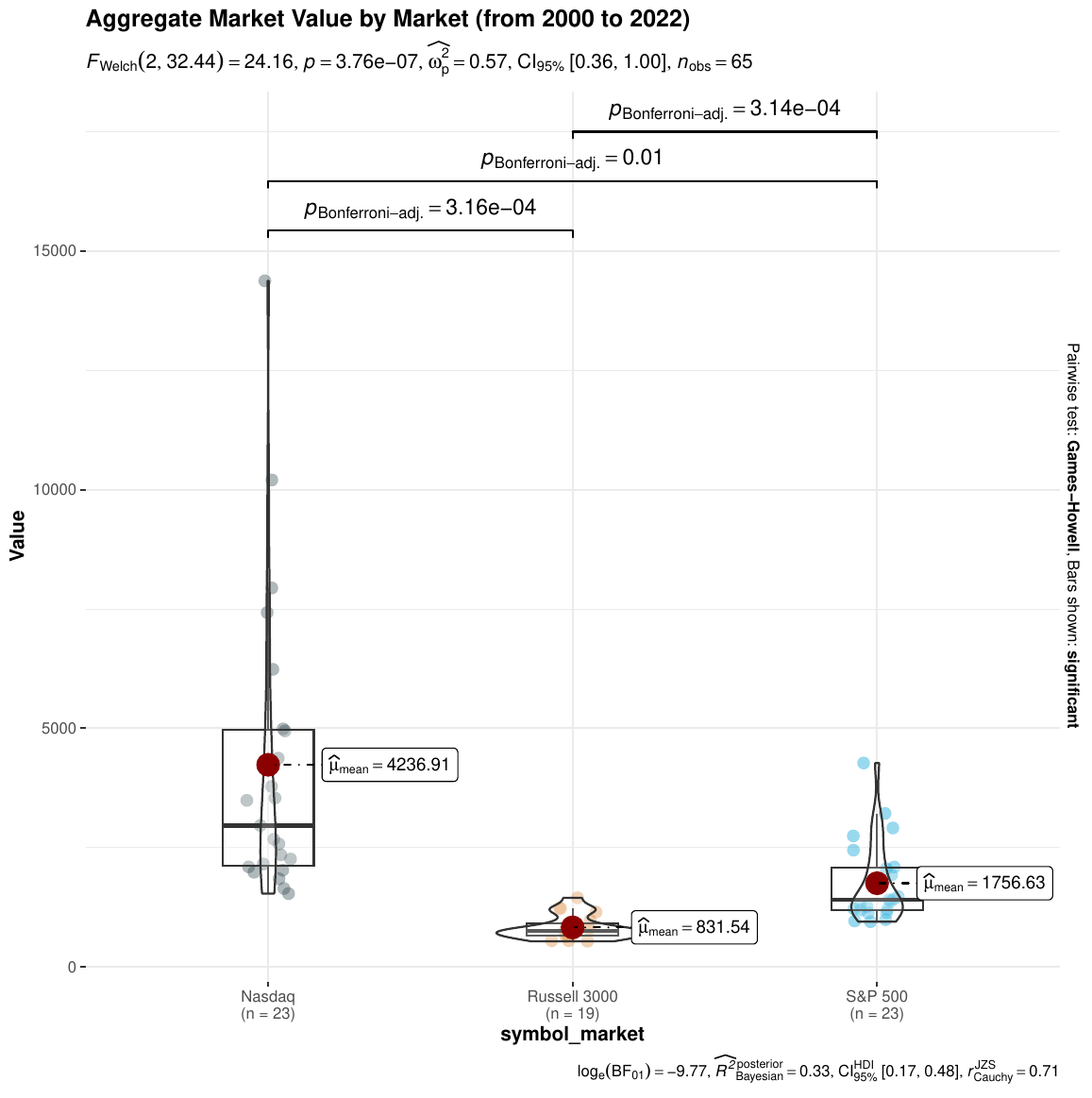}
    \caption{Aggregate value vs Market value per Market (2000-2022)}
    \label{fig:enter-label}
\end{figure*}

To construct our novel indicator, we will process a suite of technical indicators alongside historical price data, employing multiple linear regression on the clusters derived from K-Means outcomes. This step will consider the multifaceted influences of the identified variables, selecting those of greatest significance through rigorous testing.

\begin{figure*}
    \centering
    \includegraphics[width=14cm]{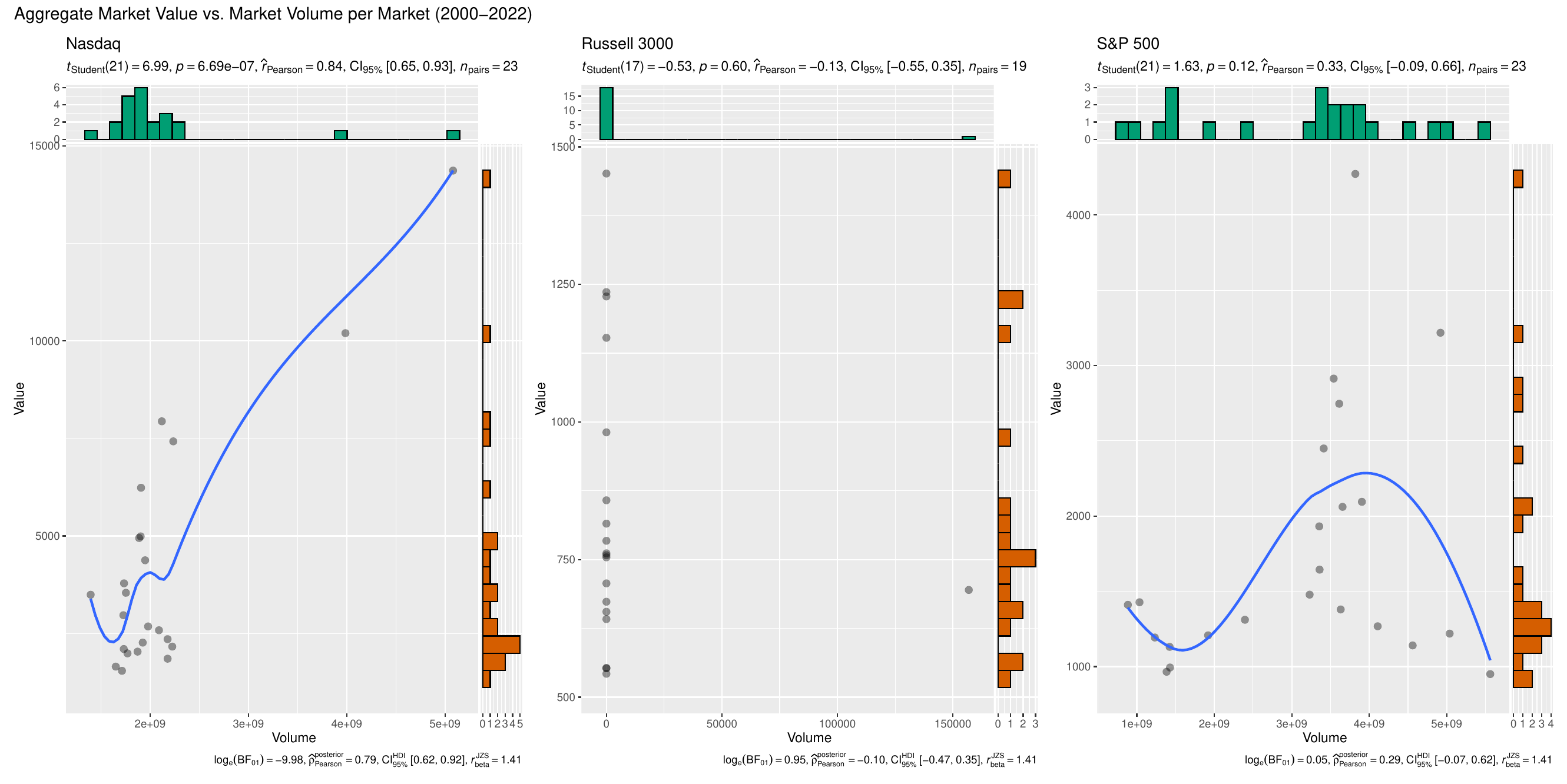}
    \caption{Correlation analysis of various factors used in this study}
    \label{fig:enter-label}
\end{figure*}

Upon developing our predictive model, we intend to classify whether forecasted price changes signify substantial and enduring shifts. Logistic regression will be utilized to formulate these predictive classifications, culminating in the creation of our new indicator. The model's efficacy will be appraised by comparing theoretical earnings achieved through adherence to the tool's guidance against baseline strategies of buy-and-hold or random trading. Model performance will be further evaluated based on R-squared and accuracy metrics, with graphical representations illustrating the evolution of indicator values over key market downturns (2000, 2008, 2020), thus showcasing trend patterns, seasonal variations, and cyclical fluctuations.

\section{Results}

\subsection{Overview}
A series of comparative analyses were conducted across different markets to ascertain the impact of each variable relative to the market type (benchmark). These analyses were instrumental in discerning the dynamics of market values over the period from 2000 to 2022, with a particular focus on NASDAQ, Russell 3000, and S\&P 500 benchmarks. Figures illustrating market value measurements employ violin and jitter plots to present a comprehensive view of market behavior over the specified period. Notably, the NASDAQ market value exhibited significant fluctuations compared to the Russell 3000 and S\&P 500, with specific outliers prominently marked by their corresponding years.

To lay the groundwork for effective predictive and regression modeling, a detailed correlation analysis was performed. This analysis aimed to delineate positive and negative correlations among the study variables, thereby facilitating an informed selection process for variables to be included in our predictive models. The correlation matrix, depicted in Figure 1, elucidates the relationships between each pair of variables, ensuring a minimally biased approach to variable selection.

\subsection{Multiple Regression Analysis}

The multiple regression analysis was an essential component of our study, designed to elucidate the relationship between various independent variables—spanning socioeconomic and financial domains—and the stock prices, our target variable. Through this analytical lens, we examined three distinct regression models, each tailored to a specific market benchmark: NASDAQ, Russell 3000, and S\&P 500. This approach enabled us to dissect the nuanced impacts of different variables across market types, thereby enhancing the robustness and applicability of our findings.

\begin{table}[!htbp] 
\centering 
\caption{Regression Analysis of Various Socioeconomic Variables with Stock Prices}
\label{tab:regressionAnalysis} 
\begin{threeparttable}
\sisetup{
    table-format = -1.4,       
    input-symbols = (),        
    group-digits = false       
}
\begin{tabular}{@{\extracolsep{4pt}} l SSS} 
    \toprule
    & {\textbf{NASDAQ}} & {\textbf{RUSSELL 3000}} & {\textbf{S\&P 500}} \\
    \midrule
    RSI & -0.293\tnote{***} & -0.320\tnote{***} & -0.317\tnote{***} \\ 
    & (0.009) & (0.009) & (0.009) \\ 
    
    MACD & 7.986\tnote{***} & 8.161\tnote{***} & 7.974\tnote{***} \\ 
    & (0.083) & (0.083) & (0.083) \\ 
    
    Market Value & 0.012\tnote{***} & 0.052\tnote{***} & 0.090\tnote{***} \\ 
    & (0.0001) & (0.0003) & (0.0005) \\ 
    
    CPI & 0.312\tnote{***} & -0.081\tnote{***} & -0.128\tnote{***} \\ 
    & (0.020) & (0.021) & (0.021) \\ 
    
    Inflation & -3.622\tnote{***} & -1.565\tnote{***} & -2.090\tnote{***} \\ 
    & (0.175) & (0.176) & (0.176) \\ 
    
    Unemployment & 0.079 & 1.514\tnote{***} & 1.666\tnote{***} \\ 
    & (0.072) & (0.073) & (0.074) \\ 
    
    GDP & -0.008\tnote{***} & -0.004\tnote{***} & -0.004\tnote{***} \\ 
    & (0.001) & (0.001) & (0.001) \\ 
    
    Yield & 0.880\tnote{***} & -0.822\tnote{***} & -0.488\tnote{***} \\ 
    & (0.090) & (0.093) & (0.093) \\ 
    
    Constant & -16.660\tnote{***} & -0.293 & 13.021\tnote{***} \\ 
    & (1.844) & (1.902) & (1.948) \\ 
    
    \midrule
    Observations & {145979} & {145979} & {145138} \\
    R-squared & {0.543} & {0.542} & {0.541} \\
    Adjusted R-squared & {0.543} & {0.542} & {0.541} \\
    \bottomrule
\end{tabular}
\begin{tablenotes}
\small
\item Note: \tnote{*} p<0.1; \tnote{**} p<0.05; \tnote{***} p<0.01
\end{tablenotes}
\end{threeparttable}
\end{table}

Our analysis revealed that the Relative Strength Index (RSI), Moving Average Convergence Divergence (MACD), Consumer Price Index (CPI), inflation rates, GDP growth, and Treasury yields exhibit significant predictive power across all examined markets. These findings underscore the multifaceted nature of stock price dynamics, where both market-specific technical indicators and broader macroeconomic factors intertwine to shape investment landscapes.

Interestingly, the unemployment rate emerged as a statistically significant predictor only within the contexts of the Russell 3000 and S\&P 500 markets. This distinction may reflect the differential sensitivity of market segments to labor market conditions, suggesting a nuanced interplay between economic fundamentals and investor sentiment across indices.

The models achieved a consistent R-squared value of 54\%, indicating that the selected variables collectively explain approximately 54\% of the variability in stock prices across the analyzed markets. This level of explanatory power highlights the complex, yet discernible, patterns governing stock price movements, and underscores the potential of our modeling approach to capture key drivers of market behavior.

During the initial phases of our analysis, the comprehensive aggregation and preprocessing of data from diverse sources laid the groundwork for our modeling efforts. An intriguing pattern emerged from this preparatory work: anomalously high stock prices were markedly more prevalent than their low counterparts. This trend, reflective of the market's inherent growth bias, hints at the prospective utility of short selling strategies under specific market conditions, further illustrating the practical implications of our analytical findings.

The nuanced insights gleaned from our multiple regression analysis not only contribute to the academic discourse on financial market prediction but also offer actionable intelligence for investors navigating the complexities of diverse market environments. As we continue to refine our models and extend our analysis, these preliminary results lay a solid foundation for developing sophisticated investment strategies informed by a deep understanding of market dynamics.

\subsection{Logistic Regression Classification}

Addressing the challenge of imbalanced data, a critical hurdle in financial market prediction, our study adopted a focused approach by selecting a single stock for model training while utilizing data from another stock to test model performance. This strategy allowed for a rigorous evaluation of the model's predictive capabilities across both familiar and unseen data scenarios. For instance, in the case of IBM, a notable imbalance was observed: among 5567 events post-cleanup and merging, only 12 were identified as drawdowns, underscoring the rarity of significant market downturns within our dataset.

Various techniques were employed to mitigate the imbalanced nature of our dataset, including down-sampling, up-sampling, and the Synthetic Minority Over-sampling Technique (ROSE). Ultimately, the up-sampling method was favored due to its superior performance in enhancing the F1 score, a critical metric for evaluating the precision and recall balance of our predictive model.

During the model-building phase, an initial set of 35 features, encompassing both raw data and interaction terms, was refined. Backward Stepwise Regression played a pivotal role in this process, enabling the reduction of feature count to 30 and subsequently to 26, following the manual elimination of features that lacked statistical significance. This meticulous feature selection process not only streamlined the model but also contributed to a reduction in the Akaike Information Criterion (AIC) values, from 357.68 to 344.85 and finally to 341.68, indicating an improved model fit.

\begin{table}[!htbp] \centering 
  \caption{Logistic Regression Results on Stock Market and Macroeconomic Indicators}
  \label{tab:logitResults}
\begin{tabular}{@{\extracolsep{5pt}}lc} 
\hline 
\hline \\[-1.8ex] 
 & \multicolumn{1}{c}{\textit{Dependent variable: Stock Market Movement}} \\ 
\cline{2-2} 
\hline \\[-1.8ex] 

\textbf{Macroeconomic Indicators} & \\
Consumer Sentiment & 0.956$^{***}$ (0.65, 1.26) \\ 
Consumer Price Index & 1.756$^{***}$ (1.19, 2.33) \\ 
Monthly Durable Goods & $-$0.0003$^{***}$ ($-$0.0004, $-$0.0002) \\ 
Fed Funds Rate & $-$21.175$^{***}$ ($-$25.79, $-$16.56) \\ 
Quarterly GDP & 0.054$^{***}$ (0.033, 0.075) \\ 
Expected Monthly Inflation & 4.657$^{***}$ (1.40, 7.92) \\ 
Monthly Retail Sales & 0.0001$^{***}$ (0.0001, 0.0002) \\ 
10-Year Yield & $-$24.390$^{***}$ ($-$37.14, $-$11.64) \\ 
5-Year Yield & 39.660$^{***}$ (26.62, 52.70) \\ 

\textbf{Market Variables} & \\
OBV & 0.00000$^{***}$ (0.000, 0.000) \\ 
MACD & 8,918.92 ($-$6,260.50, 24,098.35) \\ 
MACD Signal & $-$8,884.42 ($-$24,063.92, 6,295.09) \\ 
MACD Hist & $-$8,944.43 ($-$24,127.21, 6,238.36) \\ 
Real Lower Band & $-$21,796.90$^{***}$ ($-$32,222.07, $-$11,371.73) \\ 
Real Upper Band & $-$21,802.71$^{***}$ ($-$32,228.05, $-$11,377.38) \\ 
Real Middle Band & 43,587.53$^{***}$ (22,737.94, 64,437.11) \\ 
Open & 3.972$^{***}$ (1.36, 6.59) \\ 
High & $-$6.137$^{***}$ ($-$9.00, $-$3.27) \\ 
Close & 10.231$^{***}$ (7.84, 12.63) \\ 
Volume & 0.00000$^{***}$ (0.00000, 0.00000) \\ 
Return & $-$1,638.11$^{***}$ ($-$1,958.64, $-$1,317.58) \\ 

\hline \\[-1.8ex] 
Observations & 10,591 \\ 
Log Likelihood & $-$143.841 \\ 
Akaike Inf. Crit. & 341.683 \\ 
\hline 
\hline \\[-1.8ex] 
\textit{Note:}  & \multicolumn{1}{r}{$^{*}$p$<$0.1; $^{**}$p$<$0.05; $^{***}$p$<$0.01} \\ 
\end{tabular} 
\end{table}

The effectiveness of our logistic regression model was further assessed through the development of a Confusion Matrix, offering insights into the model's predictive accuracy by quantifying true positives, false positives, true negatives, and false negatives.

When tested against the training data, our model achieved an impressive F1 score of 0.774, signifying a robust predictive capability within familiar contexts. However, the model's performance significantly diminished when applied to previously unseen data, as evidenced by the Receiver Operating Characteristic (ROC) curve, suggesting a potential overfit to the training dataset or an inadequacy of features for generalization.

\begin{figure*}
    \centering
    \includegraphics[width=10cm]{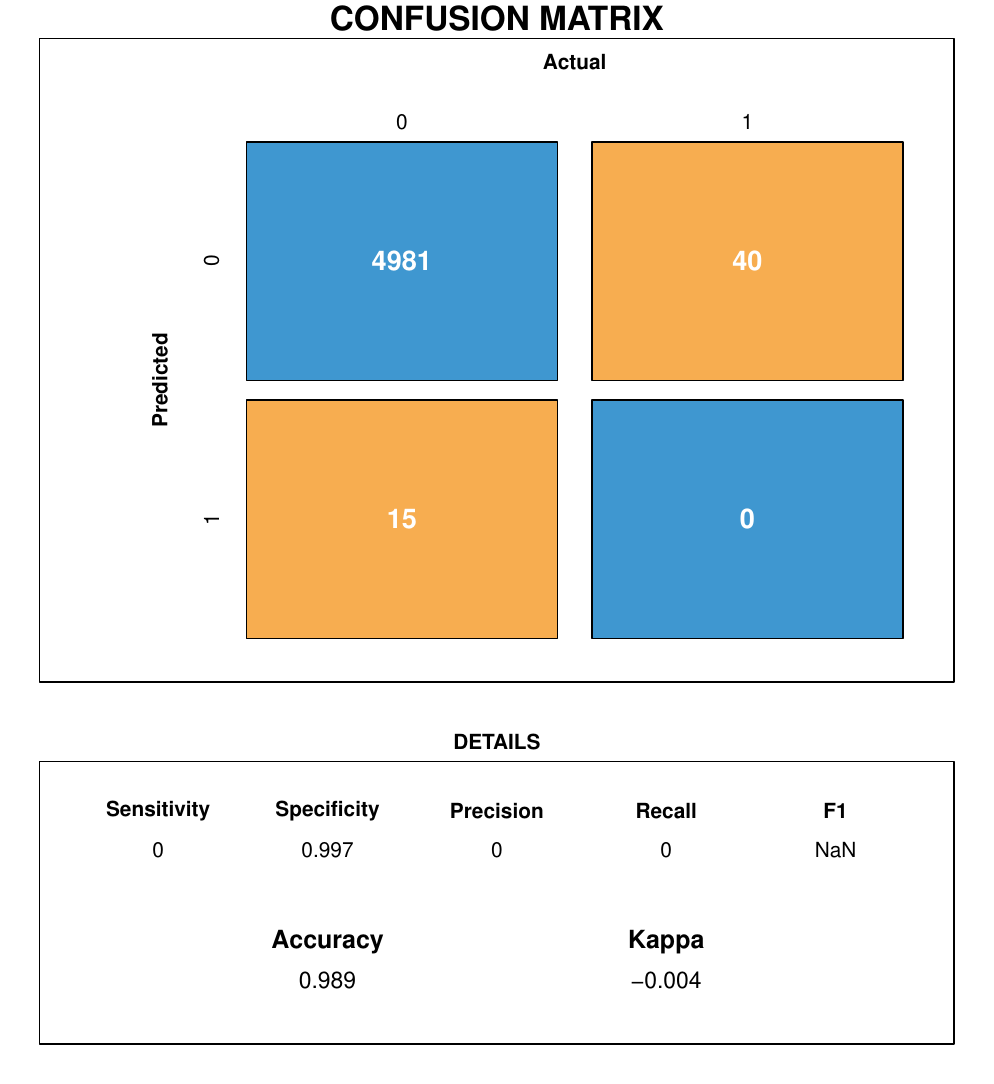}
    \caption{Confusion matrix corresponding to logistic regression classification}
    \label{fig:enter-label}
\end{figure*}

The observed overfitting and the presence of false positives raise critical considerations for market exit strategies. To enhance the model's predictive robustness and applicability, future iterations may incorporate additional forecasting methodologies, such as Holt-Winters or ARIMA models. Integrating these forecasts as either supplementary features or alternative decision-making tools could significantly augment the logistic model's accuracy, thereby facilitating more informed investment decisions.

\subsection{K-Means Clustering}

The K-Means clustering algorithm was employed to unearth patterns of similarity among the stocks within the Information Technology sector, particularly focusing on 29 selected stocks. This method aids in striking a balance between the granular, stock-specific data and the more general, aggregated market data, facilitating a nuanced analysis that accounts for individual stock performance and broader market trends.

Prior to clustering, the dataset underwent a comprehensive cleaning process to ensure data integrity and relevance. The dataset was then scaled and restricted to cover only the dates where data for all 29 stocks were available, spanning the most recent 40 years. This step was crucial to normalize the scale of stock prices and other financial indicators, preventing any disproportionate influence of larger values on the clustering process. Each stock's data was vectorized, transforming the multivariate time series into a format suitable for clustering, where each row represented a stock and each column a date-factor combination.

\begin{figure*}
    \centering
    \includegraphics[width=10cm]{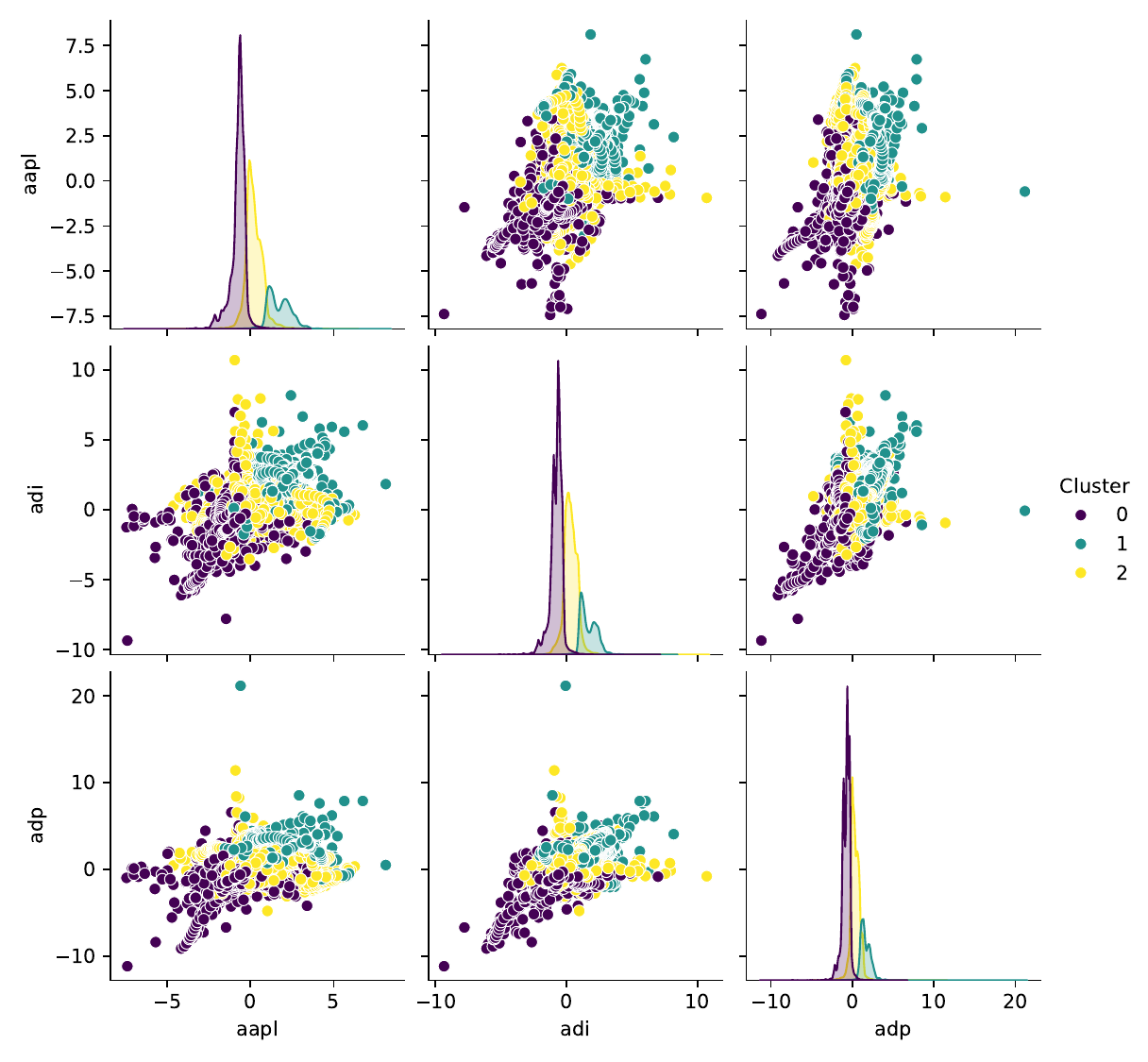}
    \caption{Pair wise clustering results visualization for AAP, ADI, and ADP companies}
    \label{fig:enter-label}
\end{figure*}

To determine the optimal number of clusters (k), we conducted ten iterations of the K-Means algorithm for each potential k value, ranging from 2 to 10. This approach was designed to mitigate the impact of random seed initialization on the clustering outcome. The analysis of these iterations involved plotting the within-cluster sum of squares against the number of clusters to identify the "elbow point," a heuristic indicating where additional clusters cease to provide substantial improvement in variance explained.

The elbow plot (Figure \ref{fig:elbowplot}) revealed that the optimal number of clusters for our dataset was three, marking a point of diminishing returns for further increasing k. This result led to the categorization of the 29 stocks into three distinct clusters, each representing a unique pattern of stock behavior over the examined period.

The clustering results indicated noteworthy groupings among the selected stocks, with some pairs of companies, such as Apple and Microsoft or IBM and Western Digital, being classified within the same cluster despite their differing corporate structures and market niches. These groupings offer an alternative perspective to traditional market segmentation and industry classification, suggesting potential similarities in stock behavior that might not be apparent through conventional analysis.

While the clusters identified through this process are not intended to prescribe definitive investment strategies, they do illuminate patterns that merit further investigation. Analyzing these clusters independently may yield insights into underlying market dynamics that influence stock performance, offering a complementary tool for portfolio management and investment analysis.

The application of K-Means clustering to financial market analysis underscores the potential of machine learning techniques to uncover latent patterns within complex datasets. Future research could explore the integration of additional variables, the application of alternative clustering algorithms, and the examination of other market sectors to broaden the understanding of stock market behavior.


\section{Discussion}

The application of machine learning models in the technical analysis of stock markets represents a significant advancement over traditional indicators. Our project endeavored to synthesize a wealth of data from technical indicators and macroeconomic variables, employing a suite of machine learning models to predict market downturns, specifically focusing on identifying optimal buying points during such downturns.

\subsection{Analytical Comparison of Model Performances}

\subsubsection{Performance Metrics}
Each model was evaluated based on its accuracy, precision, recall (sensitivity), F1 score, and the ability to generalize to unseen data. The logistic regression model exhibited commendable performance in reducing false negatives, a crucial feature that ensured investors were seldom caught off-guard by unanticipated dips. However, its precision of 63\% indicated a considerable rate of false alarms, potentially leading to missed investment opportunities. 

The random forest classifier, with its ensemble approach, offered improved accuracy and a better handle on overfitting compared to individual decision trees. Its strength lay in identifying complex patterns across a multitude of variables, making it a robust model for classifying dips versus non-corrections.

K-means clustering provided invaluable insights into the underlying structure of the market, grouping stocks with similar performance patterns. While not predictive in the traditional sense, the clusters facilitated a deeper understanding of market dynamics, guiding more nuanced analyses.

The SVM model stood out for its capacity to delineate 'buy the dip' opportunities with precision, thanks to its ability to handle high-dimensional data and find the optimal boundary between classes. Its main limitation, however, was the scarcity of data points, which constrained the model's training and validation process.

\subsubsection{Comparative Insights}
Across the board, models demonstrated distinct strengths and weaknesses contingent upon the nature of the task—whether predicting downturns, classifying market corrections, uncovering data clusters, or pinpointing investment opportunities. Logistic regression and SVM models, in particular, highlighted the trade-off between minimizing risk (avoiding false negatives) and maximizing opportunities (reducing false positives), a balance critical to investment strategies.

\subsection{Strategic Implications for Investors}

The combined use of these models offers a comprehensive toolkit for navigating market dynamics. By leveraging logistic regression and random forest models, investors can gauge the likelihood of market corrections with greater confidence. Concurrently, SVM analysis can pinpoint opportune moments for investment, further informed by the structural insights provided by K-means clustering.

\subsection{Future Directions and Improvements}

The predictive prowess of our models could be enhanced by integrating additional data sources, such as sentiment analysis from financial news and social media, which could provide early indicators of market shifts. Furthermore, employing deep learning techniques could potentially unveil more intricate patterns within the data, albeit at the cost of increased computational complexity and the need for larger datasets.

\subsection{Conclusion}

The exploration of machine learning models in predicting stock market downturns and identifying buying opportunities presents a promising avenue for both academic research and practical investment strategies. While challenges remain in model selection, data availability, and the inherent unpredictability of financial markets, our findings underscore the potential of a data-driven approach to augment traditional market analysis. Moving forward, the fusion of machine learning insights with human expertise could redefine the landscape of financial investment decision-making.

\newpage
\bibliographystyle{plainnat}
\bibliography{references}

\end{document}